# The role in the Virtual Astronomical Observatory in the era of massive data sets

G. Bruce Berriman*[a], Robert J. Hanisch[b,c], T. Joseph W. Lazio[d]

[a] Infrared Processing and Analysis Center, California Institute of Technology, Pasadena, CA 91125, USA; [b]U.S. Virtual Astronomical Observatory, 1400 16th Street NW, Suite 730, Washington, DC 20036 USA, [c] Space Telescope Science Institute, 3700 San Martin Drive, Baltimore, MD USA 21218; [d] Jet Propulsion Laboratory, California Institute of Technology, M/S 138-308, 4800 Oak Grove Dr., Pasadena, CA 91106 USA

**ABSTRACT**

The Virtual Observatory (VO) is realizing global electronic integration of astronomy data. One of the long-term goals of the U.S. VO project, the Virtual Astronomical Observatory (VAO), is development of services and protocols that respond to the growing size and complexity of astronomy data sets. This paper describes how VAO staff are active in such development efforts, especially in innovative strategies and techniques that recognize the limited operating budgets likely available to astronomers even as demand increases. The project has a program of professional outreach whereby new services and protocols are evaluated.

## 1. INTRODUCTION

Vast quantities of observational data and simulations are becoming available to astronomers at an ever-accelerating rate and are transforming astronomy, which has been a pioneer of data-driven science. New astronomical observatories anticipate delivering combined data volumes of over 100 PB by 2020 [1], yet even the current data volume of 1 PB is beginning to strain archives. At the same time, simulations are increasing in complexity and scope, approaching $10^{10}$ particles in individual simulations [2]. Concomitant with the growth in volume is the growth in the complexity of products, often derived through integrating existing data sets and confronting them with simulations by teams of distributed, often international, collaborations. Figure 1 demonstrates this growing scale and complexity of data sets.

Without intervention, astronomy will witness a breakdown in its current computing model, in which data are discovered and downloaded through web-based services offered by archives and data centers, and then analyzed and integrated on local machines. The very scale of new data sets will transform data discovery, access, and computation in astronomy. Given that maximum science return will involve federation of data sets, discovery of data or simulations will be performed through queries to distributed archives; these queries will aim to locate data having particular properties and attributes (position, wavelength time-interval, data quality…). Desktop machines will not have the power to process or even store these data, and so data will be processed in situ by archives running astronomers' software or shipped to remote processing facilities. The National Academy of Science's Decadal Survey of Astronomy and Astrophysics ("Astro2010") [4] recognized the growing impact of computation in astronomical research, and cited "cyber-discovery" as one of the necessary aspects in order to enable "transformational comprehension, i.e., discovery." Moreover, new proposals to the U.S. National Science Foundation (NSF) require the inclusion of a data management plan, whose provisions include sharing "The primary data, samples, physical collections and other supporting materials created or gathered in the course of work under NSF grants." Proposals in response to and National Aeronautics and Space Administration (NASA) Research Announcements and Cooperative Agreement Notices have similar requirements.

* gbb@ipac.caltech.edu; phone 1-626-395-1817; http://astrocompute.wordpress.com

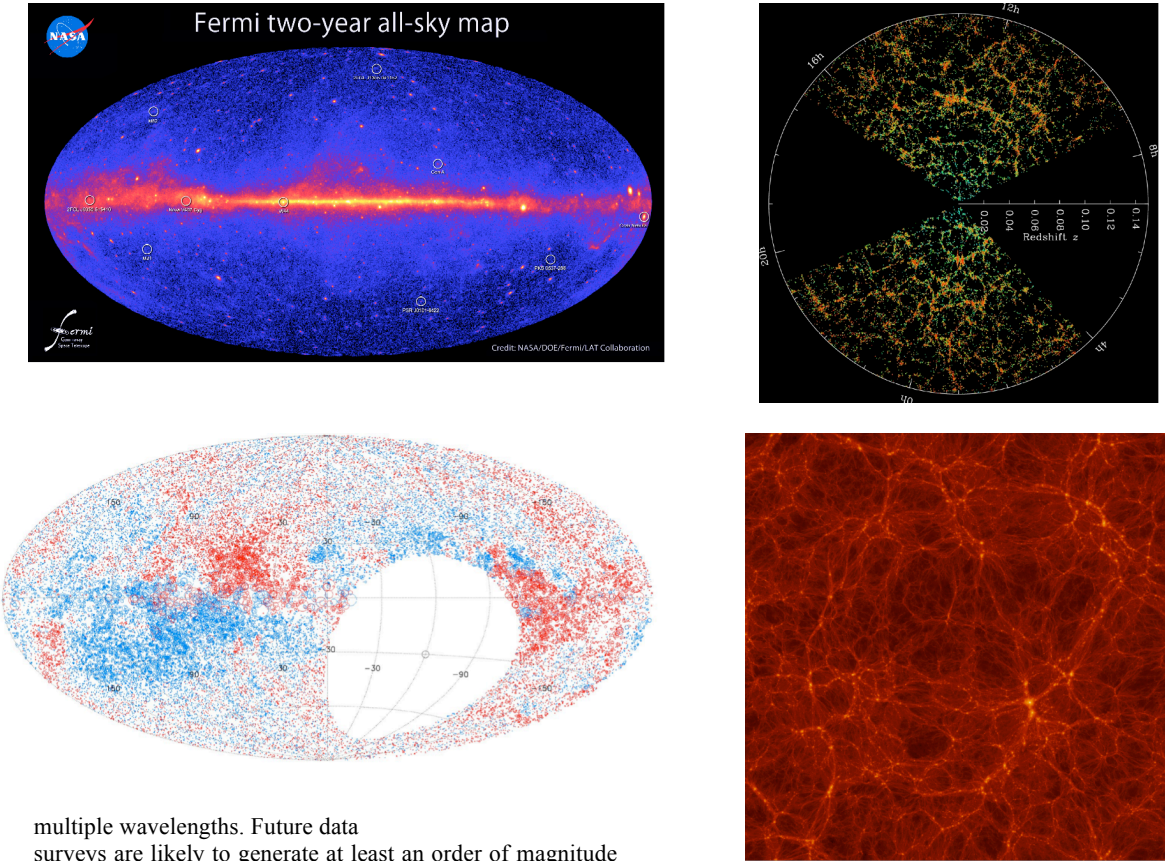

Figure 1: An illustration of the increasing scope and complexity of astronomical data sets, particularly their reach across multiple wavelengths. Future data surveys are likely to generate at least an order of magnitude more sources than these examples. (top left) Fermi gamma-ray sky, showing both identified and an increasing number of un-identified sources (Credit: NASA/DOE/Fermi LAT Collaboration); (top right) Sloan Digital Sky Survey sky; (bottom left) Subset of strongly polarized radio sources within the 1.8 million source catalog of the NRAO VLA Sky Survey ([3]); (bottom right) Slice of the 8.6 billion particle Bolshoi simulation ([2]; Credit: Stefan Gottlöber).

The "Innovations in Data Intensive Astronomy" workshop (Green Bank, WV, May 2011 [5]) described the impact of what has come to be called the "data tsunami" in all aspects of astronomy, and recognized that the need for a community effort and for partnerships with national cyber-infrastructure programs. This community effort has already begun. Next generation missions have by necessity led the way. Projects such as the Large Synoptic Survey Telescope (LSST) [6], the Square Kilometer Array (SKA) [7], the Low Frequency Array (LOFAR) [8], and others are actively investigating emerging technologies such as cloud computing and grid technologies to develop the cyber infrastructure needed to support data management and storage, and developing fast processing pipelines to enable what is often near real-time processing and dissemination of massive data streams from telescopes.

The Virtual Astronomical Observatory (VAO; http://usvao.org) anticipates playing a major role in this community wide endeavor. The VAO has been in operation since May 2010, funded jointly by the NSF and NASA, for the express purpose of developing an information backbone for the astronomical research community that responds to the growing scale and complexity of astronomical data. In its first year, it delivered four new services aimed at understanding the needs of the backbone: a Spectral Energy Distribution builder (SED), Iris; a catalog cross-comparison engine; a data discovery service and a time-series discovery and analysis service. In addition, multi-archive VO-compliant data access services are now incorporated into the Image Reduction and Analysis Facility (IRAF).

The VAO participates in the development of standardized data discovery and access protocols and data description models as a member of the International Virtual Observatory Alliance (IVOA). This paper will not describe these proposals and models in detail other than to recognize that they will remain central to the discovery of data and will continue to underpin the information backbone. Rather, we focus on progress on backbone itself, which will take several more years to complete. New projects are cooperating with the VAO to support the development of this backbone to serve and distribute their datasets and take advantage of re-usable tools that allow their end users to discover and explore new data sets, with obvious costs benefits to the projects and obvious technical benefits to users who can exploit a common set of tools for may purposes. The paper describes developments to date and future plans for deploying a complete environment, shown in Figure 2.

This is one of a series of papers in the SPIE "Astronomical Telescopes and Instrumentation 2012" conference that describes the work of the VAO. Companion papers describe the organization and management of the VAO [9], operations [10], management of distributed software [11] and support for time series astronomy [12]. The last paper and a companion paper [13] discuss the application of the information backbone to time-series data sets, so this topic will not be discussed in detail here.

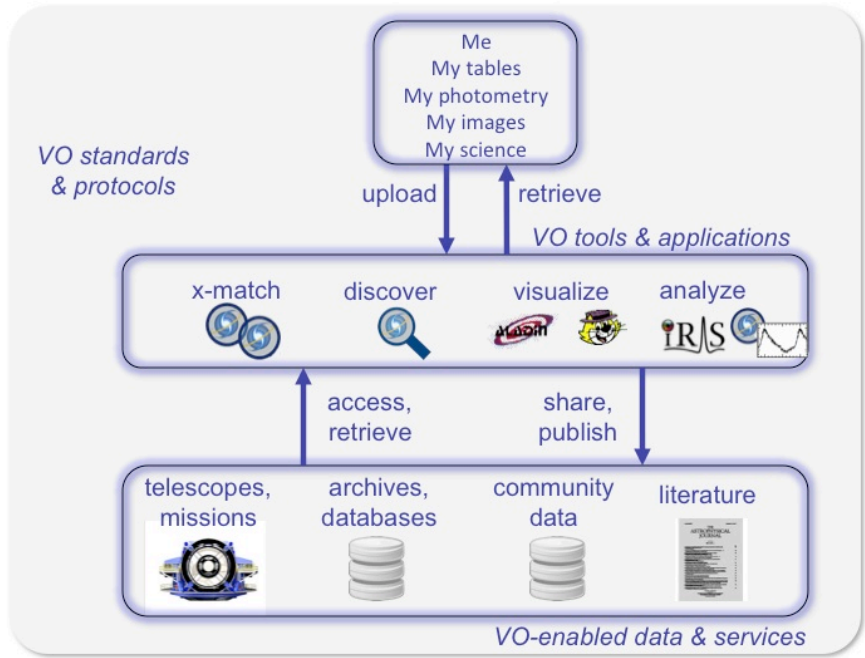

Figure 2: The Virtual Observatory will, when fully developed, provides an environment in which tools and applications can interact seamlessly with diverse data collections, including telescope and mission archives and databases, data published with the literature, and with a distributed repository for community-generated data. International standards and protocols for data discovery and access underlie this "ecosystem". VAO applications for cross-matching, data discovery, spectral energy distribution analysis, and time series analysis, work in concert with general VO tools such as Aladin (image display and visualization) and TOPCAT (tabular data display and manipulation). Users can build their own tools to plug into the ecosystem, and can choose to share data with collaborators or publish data to the community at large.

## 2. COMPUTATIONAL INFRASTRUCTURE FOR ASTRONOMY: THE VAO IN CONTEXT

The increase in data volumes, the growth in the number of products and the growing importance of archival research in astronomy are already affecting the performance of archives and data centers. Many operate on budgets that are fixed for several years, and cannot correct degradation in performance simply by adding hardware infrastructure as usage increases, as is common in commercial enterprises. This degradation is, for example, already affecting the NASA Infrared Science Archive. It is going through a period of exceptional growth because it is assuming responsibility for the massive data sets released by the Spitzer Space Telescope and the Wide-field Infrared Survey Explorer (WISE) mission. The volumes of these two data sets alone exceed the total volume of the missions and projects already archived, and the response times to queries have already suffered, primarily because of a growth in requests for large volumes of data [14]. Moreover, the Spitzer data alone includes over 20 complex, contributed data sets that usually integrated Spitzer data with data sets having broad wavelength coverage. A second example of load is in the area of distribution and management of transient event notices. The LSST anticipates distributing as many as 2,000,000 notices per night, but other telescopes are already producing event streams of sufficient quantity that a robust system must be maintained and expanded soon (see [12] and [13])

The astronomy community thus an immediate need for innovative approaches in discovering and accessing data sets that respond to the growth in usage yet without putting excessive load on archive servers. Such services will be a major part of the information backbone that will be developed by the VAO. Section 3 describes developments to date and summarizes future approaches,

New surveys and instruments that are in some cases close to their commissioning phase have begun to take up VO services as they begin to develop their data management and archive systems Developing the infrastructure to handle the processing and storage needs are major priorities for these projects, and they are looking towards off-the shelf solutions to support the development of data discovery, data access and data exploration services, and to support the development of data exploration by end users. The Dark Energy Survey (DES), which will use a custom camera to probe why the Universe is accelerating, and plans to use VO standards to serve images and source catalogs. Indeed, the project has already successfully demonstrated the use of the VO standard for access to catalogs, the Table Access Protocol (TAP), through a dedicated client called "Seleste." The One – Degree Imager (ODI) is facility instrument at NAOA that will capture one-square degree images of the sky at optical wavelengths. Its Pipeline, Portal and Archive (PPA) system has already built in the use of VOTable, the VO standard for transferring data, encoded in XML. Future releases plan to use VO image and catalog standard. The Atacama Large Millimeter Array (ALMA) plans to take advantage of Release 2 of the VO image protocol, Simple Image Access Protocol (SIAP), to serve image cubes, the project's primary data product. Future plans include exploiting VOSpace, and interface to manage distributed storage space. Accordingly, the VAO has begun to establish formal collaborations with these teams to ensure that VO services fully meet their needs and, of course, those of their user communities.

Projects that are still in the planning stages are by necessity investigating and exploiting emerging technologies to provide cost-effective infrastructure for processing and for data storage. The Australian SKA Consortium is proposing to use a cloud solution based on machines in the Australian academic network instead of expensive commercial clouds [15], while the LSST is building a high-performance infrastructure at the National Center for Supercomputing Applications (NCSA) [16]. Projects such as these also expect to exploit VO services to provide cost-effective data access and data exploration services, Section 4 describes these services and how they can be re-used to create what might be called an "ecosystem, of tools."

VAO team members are in many cases members of these project teams and are supporting the development of the infrastructure for large projects and have expertise in data management. The VAO is taking advantage of this knowledge in advising the community in data management and data sharing. Team members are, for example, advising the NASA Exoplanet Science Institute's (NExScI) Sagan Workshop (July 2012) in using cloud computing to support hands-on science analysis sessions. These sessions will involve 120 participants simultaneously analyzing exo-planet light-curves.

Finally, teams of astronomers are starting to integrate archived data with new observations, often as part of large teams, to create yet more large data sets. There is a growing need for distributed disk space to house these data, in preparation for dissemination to the community, but archives do not have the resources to provide, organize and manage such space on behalf of these teams. In response, the VAO has begun a major initiative, described in Section 5, to support data sharing and data publication.

## 3. DATA DISCOVERY AND DATA ACCESS

To maximize the science return from telescopes, it is necessary to access data to create science projects that may not have been envisioned at the time that the data were acquired originally ("data re-purposing"). As an illustration of the potential power of data re-purposing, the number of papers written by accessing data from the archives at the Space Telescope Science Institute [17]. Efficient data discovery is therefore of crucial importance, given the growth in the number and complexity of data sets.

The VO has pioneered the discovery of data across multiple archives. Applications such as Datascope (developed by the National Virtual Observatory, the predecessor of the VAO) respond to positional searches to discover data across archives distributed worldwide, including the Centre de Donnees de Strasbourg (CDS), the NASA Extragalactic Database (NED), and the NASA wavelength-based archives. Datascope queries these archives and returns the data to the user's desktop. It returns links to previews of individual data sets as well as the metadata describing the data, summaries of queries that returned no data and reports on services that did not responding.

Generally speaking, data discovery services perform these operations by taking advantage of standardized services that are recorded in a Registry. The standardized services follow a set of rules ("VO protocols') for describing queries and for encoding the response in a self-describing. eXtended Markup Language (XML) format called VOTable that can read by any computer. Applications contact the Registry, find all the data access services that are collected and organized in it, and then automatically send queries to distributed archives and organize the results for presentation the user.

By design, early experiments with publishing data to the Registry and VO standards emphasized simplicity, with minimal requirements, and led to a rapid take-up of VO publishing and demonstrated the value of standardized services in finding and accessing data worldwide: without these standards, cross-archive queries would have to be composed according to rules developed by individual archives. These experiments were also pathfinders in understanding how to develop an operational registry, given that the astronomical community lacked experience and expertise in this area. Thus, resource descriptions are very uneven in the quality of the metadata; many descriptions lack key terms that capture the essence of the underlying data collection and which would allow them be connected with a scientist's questions. The registered resources are a mixture of original and replicated collections, surveys, pointed observations, and derived catalogs. Frequently the same collection provides multiple service and some collections are accessible only through non-VO-standard interfaces The Registry does not provide enough intelligence as to the relationships between all these resources and services to adequately inform the user about what they have found. Consequently, the user can be easily confused by large search results. There are varying levels of compliance with IVOA standards and even varying levels of upkeep of not only the resource metadata but of the underlying services themselves, meaning that the registry needs much more dynamic flagging of resource quality into the registry. Finally, the Registry captures information at the data collection and service level, while ultimately users want to drill down to individual datasets and catalog records. Consequently, many of the kinds of discovery users want to do—such as, "find me all data about this source"—result in data queries going out to all known archives and repositories.

Based on this experience, the VAO plans to redevelop the VO Registry so that its scales up in several dimensions: number of datasets, area of sky, and types of datasets.  This ambitious plan will take several years to develop, but will build on prior work and will involve the following developments:  (1) a non-centralized framework for gathering metadata about available data and services, (2) an extensible metadata model and interchange format, (3) an established culture in which providers are willing to register their assets through easy-to-use registration methods, (4) a base of applications and users that use a registry for discovery along with experience with how they want to use a Registry, and (5) a method of indexing datasets to enable rapid discovery of large data sets over large spatial areas. To bring on this new era, we will take a comprehensive approach that seeks to connect publishing, discovery, and cooperative (potentially even crowd-sourced) curation of metadata and services. We aim to create a comprehensive publishing and curation system that (1) redefines the best practices for publishing to the VO that optimizes understanding by the scientist looking for data, (2) re-organizes existing resources according to best practices, and  (3) creates a system for resource description annotations that allow Registry maintainers (namely the VAO Registry maintainers) to augment and organize the view of available collections to make them more understandable by users.

Work has already begun in these areas. A new, simple interface for publishing assets has been developed and is now under test, with an anticipated delivery in Summer 2012. The VAO has begun to investigate indexing schemes that will support fast, scalable access to massive databases and will avoid placing excessive loads on archive servers. It has developed an early release of an R-tree based indexing scheme in which the indices are stored outside the database, in memory-mapped files that reside on a dedicated Linux cluster.  It offers speed-ups of up to 1000× over database table scans, and has been implemented on databases containing 2 billion records and TB-scale image sets. An early version of this is in operational use in the Spitzer Space Telescope Heritage Archive and the WISE archive [14].

In addition, a catalog cross-comparison engine, used in the VAO catalog cross-match service, takes advantage of custom indexing techniques that can be stored as files and integrated with the registry. The engine finds all those records for sources in an on-line catalog that represent candidates for identification with the sources in a users' catalog.  By design, the engine only performs this cross-matching of catalogs, but does not perform any identification of sources. In the long term, the VAO expects to integrate cross-identification schemes developed within the community, rather than develop them internally. Astronomical source catalogs usually employ spatial indexing schemes based on sky tessellation schemes such as the Hierarchical Triangular Mesh and HEALPix. They are usually stored as B-trees, with a single area of the sky having a single identifier. Bulk cross matches between very large source catalogs of the 1 billion records scale are done with more specialized "zoning" algorithms.  The data for each catalog are organized into declination strips a few arc minutes wide, with the optimum size having been determined largely by trial and error.  Inside each strip, the data are ordered by Right Ascension and the algorithm keeps three strips in memory at once (to allow for neighbors next to each other across strip boundaries). Then the cross-match engine walks through the center strip for one of them and follows along in the other, checking potential matches. Once that strip is done, we shift up and read in the next strip from each. The algorithm matches a one million record input catalog with the Sloan Digital Sky Survey (SDSS) in 275 s, and the algorithm could be parallelized to improve performance. The drawback of this scheme is not in the performance of the engine, but in the overhead in deriving the indices, which can take several days for catalogs containing a billion or so catalogs.  Cross-matching for dynamic catalogs will need further development if current practices prove inadequate.

## 4. AN ECOSYSTEM OF USER-FRIENDLY TOOLS

One of the major goals of the VAO is to enable new science by providing fast scalable access to data sets worldwide and to support users' workflows as they explore and analyze these data.  The approach is to bring the VAO to astronomers, rather than expect astronomers to learn a new way of doing business.  There are three prongs in this approach, and they will be discussed in turn:

- Integration of VO capabilities into widely used tools
- Development of capabilities that abstract the VO computational architecture from the end user
- Development of tools that are re-usable and can be plugged together to provide new capabilities

**4.1 Integrate VO capabilities into widely used tools**

The astronomical research community utilizes a variety of desktop computer environments and tools for data reduction and analysis. There are large, general purpose systems that originated from and are supported by organizations within astronomy, such as the Image Reduction and Analysis Facility (IRAF), Astronomical Image Processing System (AIPS), Common Astronomy Software Applications (CASA), Chandra Interactive Analysis of Observations (CIAO), and ESO-MIDAS.  Commercial packages, most notably IDL, and public domain environments like Python and R, that have wide or growing use, owing to the ease of prototyping and development. (Python is also becoming an integral part of environments such as IRAF and CASA.) And there are general purpose environments such as IDL and MATLAB, as well as many special-purpose applications, such as DAOPhot and SExtractor for photometry, that are in wide use.

The VAO plans to bring VO-based capabilities into the environments and applications that astronomers already know. This long-term effort will build on the integration of VO capabilities into IRAF, recently released as IRAF Version 2.16, on March 23, 2012. IRAF users may now access data from the network via arbitrary URLs and read data in VO formats (like VOTable). In addition, there is a new VO module that allows a user to find data and download them into the desktop IRAF environment.  Another important feature is the ability to use the Simple Access Messaging Protocol (SAMP; [18]) to exchange data and commands with other VO applications, creating an integrated suite of tools more powerful than the individual components. IRAF users may now, for example, send images for visualization in Aladin. IRAF is now able to query/access remote data using standard HTTP protocols and make use of the XML documents and FITS files returned using native interfaces. Extensions to the scripting environment allow for creation of new science tasks based on these enhanced capabilities. In the long term, VO/IRAF concept will be expanded through collaborations with the organizations that support CASA (NRAO), CIAO (SAO), and PyRAF (STScI).

### 4.2 Development of capabilities that abstract the VO computational architecture from the end user

VO standards offer increasingly powerful and sophisticated capabilities for end users, but at the same time the standards themselves become increasingly complex. The VAO will provide tools that make it easy for end users to take advantage of the capabilities. An example is services that use the TAP protocol, which will provide users a common and quite general view of catalogs of all sizes. In the first instance, the VAO is developing a graphical tool for creating and submitting sophisticated queries to any compliant TAP service. In particular, it allows a user unfamiliar with ADQL and TAP to connect to a TAP service, browse its database schema, construct and submit ADQL queries, and visualize the results.

### 4.3 Development of tools that are re-usable and can be plugged together to provide new capabilities

The astronomical research community has already begun to take advantage of SAMP as a means of plugging together existing tools to create powerful new analysis environments.  The PanSTARRS science applications team recently developed "PanSTARRS VO," a Java-based interface to the PanSTARRS database that interacts seamlessly with VO tools such as Aladin [19] and TOPCAT, a catalog visualization and analysis tool [20].  Figure 3 shows this toolkit in action.  The Dark Energy Survey project has utilized the TAP service implementation to expose their database to the Seleste TAP client, and query results are then easily examined with TOPCAT and Aladin..  With technical advice from VAO, Penn State's Center for Astrostatistics released VOStat Rev. 2..  Users can seamlessly access their VO data products through SAMP, and interactively apply ~40 statistical functions from the public-domain R statistical software environment [21] . These tools include data smoothing, function fitting with goodness-of-fit tests, sample comparisons, bivariate and multivariate regression, spatial analysis, non-detections, time series analysis, directional statistics, and a variety of graphical displays. VOStat Rev. 2 also provides R scripts and help files giving the user a start on a more complete statistical analysis of their data on their home computers.

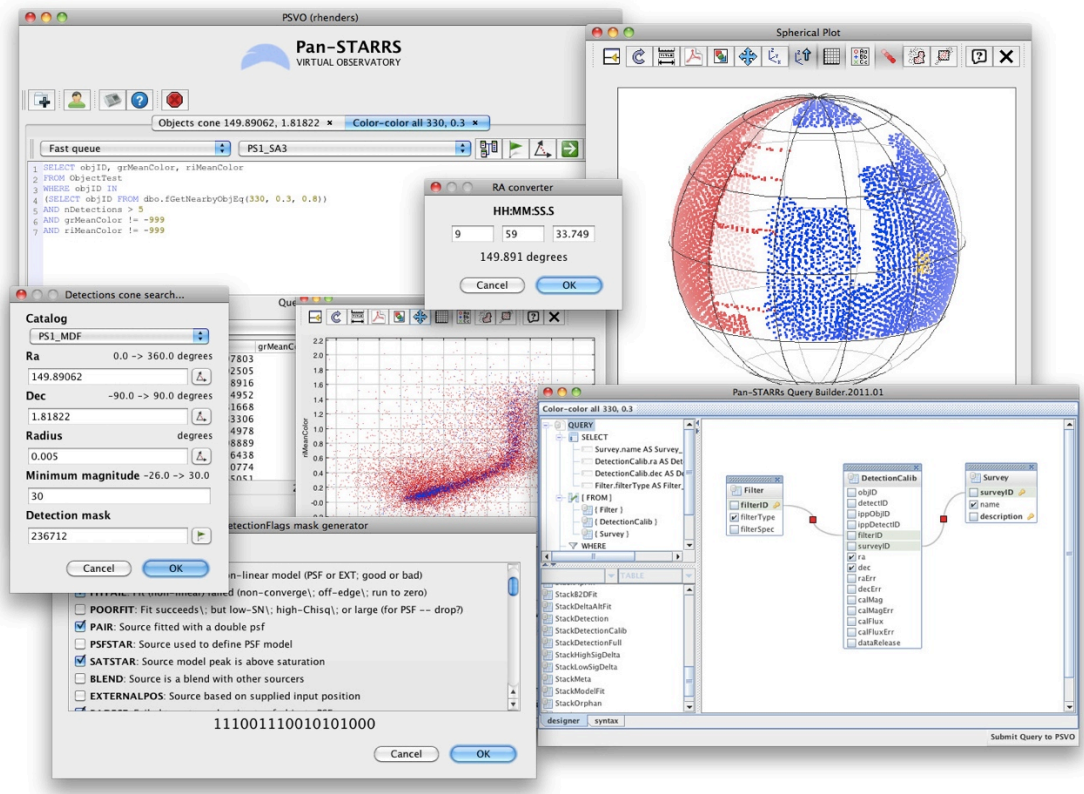

Figure 3: The PanSTARRS VO environment (courtesy R. Henderson, U. Hawaii)

A major initiative for the VAO is to "SAMP-enable" all its existing tools so that the ecosystem of tools can be expanded. A user of the cross-match engine would, for example, send the output records directly to TOPCAT for analysis, and would explore the cross-match candidates in more detail by finding images at multiple wavelengths that were delivered by survey projects.

## 5. THE VAO DATA SHARING AND PUBLICATION INITIATIVE

All proposals submitted after January 18, 2011 to the United States' NSF now require the inclusion of a Data Management Plan, whose provisions include sharing "The primary data, samples, physical collections and other supporting materials created or gathered in the course of work under NSF grants." Proposals in response to NASA Research Announcements and Cooperative Agreement Notices have similar requirements, "To facilitate data sharing where appropriate, as part of their technical proposal, the Proposer shall provide a data-sharing plan and shall provide evidence (if any) of any past data-sharing practices." Considerable value has resulted from being able to re-use or re-purpose data in Observatory Archives to create new products, usually through integrating several data sets. The VAO has therefore just begun a major initiative in this year to support sharing and publication of community-generated data. This initiative will provide the community with the tools and methodologies for sharing and publishing their data, organized and fully annotated to allow for discovery and maximal scientific exploitation, and will act as consultants with teams of astronomers on best practices in curating and serving data.

The data sharing services developed by the VAO will of course be open to the entire astronomical community. The primary customers for VAO data sharing services are most likely to be teams of astronomers, quite probably geographically dispersed, preparing new data sets in support of research projects. These data sets will almost certainly include proprietary data from recent observations that providers will make public, models and simulations, or new datasets derived by integrating and synthesizing extant data (e.g., customized catalogs). While the data may ultimately reside in thematic or observatory archives, the archives themselves generally do not have the resources to support sharing of what are usually highly processed data sets, that are increasingly TB-sized.

The VAO will bring to bear the expertise gained in curating and describing scientific data sets produced by large missions and projects and will as far as possible take advantage of existing technologies. The following components are essential parts of the system:

- Organized, secure storage space for data sharing and long-term data preservation, and mechanisms for delivering data. This storage space will take advantage of VOSpace, an interface for managing, caching and sharing files in a distributed storage space. In the first instance, disk space will be made available at Johns Hopkins University (JHU) and the Infrared Processing and Analysis Center (IPAC), and we are negotiating with the San Diego Supercomputer Center (SDSC) to have space made available there too.
- Tools for assessing the completeness of descriptions of data sets (metadata), for assessing compliance with astronomical data formats, for repairing defective or incomplete data sets, and for extracting metadata. There are tools available in the community for performing these tasks. The VAO act as a "clearing house" for these tools, and will augment them or develop new tools per requests from the community.
- Tools for publication of data to the VAO, so that data sets can be discovered, explored and analyzed with the VAO "ecosystem of tools." This will support the generation of new, highly processed data sets as well as validation and exploration of the shared data sets.

When complete, these components will be part of the overall VAO architecture for publishing, discovering and accessing data, as shown earlier in Figure 2, and available to all astronomers as needed. Those developing new products and wishing to store data locally may, for example, only wish to take advantage of the metadata and data validation tools

As the first step on developing the full suite of capabilities outlined above, the VAO is performing pilot studies in collaboration with research teams who have agreed to be early adopters. The goal of the pilot studies will be to inform specifications for the complete development of tools beyond the next year. The pilot studies are:

- **Data sharing:** collaborate with CANDELS, a Hubble Legacy program, in sharing data products that are under development.
- **Data publication:** collaborate with other community research groups who have data products that they wish to publish through the VO, such as astrometry.net and American Association of Variable Star Observers (AAVSO). In particular, the VAO recently accepted, in response to the VAO's call for Collaboration proposals, a proposal from the AAVSO specifically to make their data more widely available to the community.

## 6. CONCLUSIONS

This paper has described the role that the VAO has begun to play in the development of an information backbone that will support discovery, access and exploration of large and complex distributed data sets, and outlined its plans for future development. Even the current volume of data and numbers of data sets accessible through archives is degrading service quality and hampering data discovery. Thus, the VAO is active in developing innovative new ways of discovering and accessing data that aim to avoid placing excessive loads on archive servers.. The VAO is actively working with projects and teams to develop services and tools that will underpin their data access and data exploration services, and will provide simple mechanisms for sharing data.

## ACKNOWLEDGMENTS

This paper describes work done with the support of the US Virtual Astronomical Observatory. The VAO is jointly funded by the National Science Foundation (under Cooperative Agreement AST-0834235) and by the National Aeronautics and Space Administration. The VAO is managed by the VAO, LLC, a non-profit 501(c)(3) organization